# The Effectiveness of Traditional Tools and Computer-Aided Technologies for Health and Safety Training in the Construction Sector: A Systematic Review

Yifan Gao[1], Vicente A. González[2], and Tak Wing Yiu[3]

**Abstract:** For workers, the exposure to on-site hazards can result in fatalities and serious injuries. To improve safety outcomes, different approaches have been implemented for health and safety training in the construction sector, such as traditional tools and computer-aided technologies (e.g., serious games and virtual reality). However, the effectiveness of these approaches has been barely explored. In order to bridge this gap, a systematic review of existing studies was conducted. Unlike previous review studies in this field that focused on uncovering the technology characters and challenges, this study mainly evaluated the effectiveness of training using traditional tools and computer-aided technologies on the well-being of individuals. Measures of the effectiveness included knowledge acquisition, unsafe behaviour alteration, and injury rate reduction. Results indicated that: 1. the effectiveness of traditional tools is sufficiently supported by statistical evidence; and 2. the use of computer-aided technologies has evidence to support its effectiveness, but more solid evidence is required to support this statement. It was also found that the overall performance of computer-aided technologies is superior in several technical aspects compared to traditional tools, namely, representing actual workplace situations, providing text-free interfaces, having better user engagement, and being more cost-efficient. Finally, using the systematic review findings, a theoretical framework is proposed as a potential solution to help future research in this field systematically examine the effectiveness and usability of their approaches. This framework is theoretical in nature and requires further validation. A further study is therefore proposed to test and validate this framework.

[1] PhD. Candidate, Dept. of Civil and Environmental Engineering, Faculty of

Engineering, Univ. of Auckland, New Zealand, ygao516@aucklanduni.ac.nz

[2] Senior Lecturer, Director of Advanced Computing and Virtual Technologies in Construction (ACVTC) Research Group, Dept. of Civil and Environmental Engineering, Faculty of Engineering, Univ. of Auckland, New Zealand, v.gonzalez@auckland.ac.nz

[3] Senior Lecturer, Dept. of Civil and Environmental Engineering, Faculty of Engineering, Univ. of Auckland, New Zealand, k.yiu@auckland.ac.nz

# 1. Introduction

## 1.1 Health and Safety Situations

The unprecedented scale of building and infrastructure development is driving the demand for more resources and intensive use of labour. As a result, the construction sector plays a pivotal role in providing capital resources for this development and the resulting economic growth. This sector is the fifth largest industry in the New Zealand (NZ) economy, generating more than 6% of overall GDP (MBIE, 2013). This sector employed 36,000 more workers in 2012 than it did in 2002, indicating a 30% increase in the construction workforce (NZ.Stat, 2013).

Workplace accidents in this sector are also on the rise, resulting in higher rates of injury, decreased productivity, and increased compensation demands (Zhao, McCoy, Kleiner, Smith-Jackson, & Liu, 2016). In NZ, it was estimated that about 15% of all occupational injuries generated in 2015 occurred in the construction sector, causing a substantial economic loss of roughly $108 million (ACC, 2016). In China, the construction workforce had the third highest rate of occupational fatalities on record, with the number of claims increasing at a constant rate from 1280 in 1997 to 2197 in 2014 (SAWS, 2015). In the United Kingdom, the construction sector represented only

5% of the British workforce, but it accounted for almost 30% of overall occupational fatalities in 2015 across all industry sectors (HSE, 2016). It is clear therefore that the global construction sector is now facing unacceptable levels of health and safety (H&S) risks.

## 1.2 Tools for Health and Safety Training: A Sketch

H&S training has been essential to the success of construction projects (Guo, Yu, & Skitmore, 2017). "It is critical that construction personnel are trained to identify hazards on sites and to react appropriately to control those hazards", (Greuter & Tepe, 2013, p. 2). The inadequate training of frontline workers has been identified as one of the major causes of accidents in the construction sector (Choudhry & Fang, 2008; Li, Chan, & Skitmore, 2012). Further evidence appears in Council of Labour Affairs' (2007) statistics on occupational safety in Taiwan, which indicated that 67.82% of all injured workers in the construction sector did not receive safety training.

There are a number of ways in which H&S training can be implemented, such as traditional tools (TT) and computer-aided technologies (CAT). TT includes traditional techniques such as toolbox talk, video demonstration, text-based handout, hands-on training, computer-based instruction, and lecture session. The term engagement, which refers to the individuals' interests and attention in relation to the training contents during a learning process, is adopted to differentiate TT approaches: least engaging (lecture, video demonstration, toolbox talk, and text-based handout), moderately engaging (computer-based instruction), and most engaging (hands-on training) (Burke, Sarpy, Smith-Crowe, Chan-Serafin, Salvador, & Islam, 2006). However, TT has been considered by some researchers as not an ideal solution for the construction workforce (Guo, Li, Chan, & Skitmore, 2012; Li et al., 2012). As suggested by Nielsen (2015), the effective transference of H&S knowledge needs to be in accordance with employees' preferences. Construction workers have been identified as experiential learners who tend to lose interest in memorising safety regulations, lack continuous engagement with TT approaches, and would prefer more proactive learning styles (Harfield, Panko, Davies, & Kenley, 2007).

The criticisms of TT have led to an intensive search for new methods, such as computer-aided technologies (CAT). Latest innovations in CAT, such as Computer-generated Simulations (CGS), Serious Games (SG), and Virtual Reality (VR) have now matured. These technologies are nowadays standing out as interactive and portable solutions to support H&S training (Xie, Tudoreanu, & Shi, 2006). CGS can be defined as the traditional screen-based simulation that aims to provide virtual but meaningful scenes to describe the dynamic nature of construction sites (Bosché, Abdel-Wahab, & Carozza, 2015). SG is a type of screen-based video game that explicitly involves educational contents and constructs to address real-life issues (Harteveld, 2011). The aim of SG is to make players face challenges that make sense to real-life situations in order to teach specific topical areas or knowledge and provide tangible learning outcomes (Gao, Gonzalez, & Yiu, 2017). The process of designing a SG requires the modelling of interactive environments and the creation of consistent storylines (Williams-Bell, Kapralos, Hogue, Murphy, & Weckman, 2015). SG can be better understood by using the three-world theory, where a SG is composed of the worlds of meaning, reality, and play (Harteveld, 2011). The *world of meaning* is the most important factor, which considers the topical area and knowledge to be included for end-users (Harteveld, Guimarães, Mayer, & Bidarra, 2010). The *world of reality* relates to the realism that a game environment reaches in comparison to real-life environments (Harteveld, 2011). The *world of play* involves all the strategies employed in the game to sufficiently engage players so that they complete the entire game (Harteveld et al., 2010). *VR* can be defined as a computer technology enabled by VR headsets in which the realism of its display is close to real life and participants are fully immersed by visual impact to shape illusive feelings of physically existing in the virtual environment presented (Lavalle, 2015). In contrast to the screen-based CGS and SG, it is distinctive of VR to provide users with an experience that is indeed very close to what the real world looks like (Lavalle, 2015). Users move their heads towards visual cues that are not in front of them but cover a much wider arc of the vision angle (Winn, 1993).

## 1.3 Research Objectives

As already mentioned, TT approaches have been recognised as not ideal for the construction sector, and the use of CAT is in a position to substitute TT. However, this is an assumption, so the systematic review outlined here aims to identify pertinent evidence, discuss the effectiveness of implementing TT and CAT for H&S training in the construction sector, and come to a conclusion regarding the findings.

This paper is organised as follows: The methodology for this review is described in Section 2. This is followed in Section 3 with the interpretation of evidence from the literature. Based on technical aspects, Section 4 discusses the review findings and compares the uses of TT and CAT. Finally, Section 5 concludes with the main findings of this review.

## 2. Method

In order to meet the research objectives, a systematic review was undertaken in March 2018 based on the guideline proposed by Khan, Kunz, Kleijnen, and Antes (2003).

In step one, research questions for this review were framed: 1. is there enough evidence available to justify the premise that training using TT produces a positive change in behaviour, knowledge or injury rate? 2. is there enough evidence available to justify that training using CAT (i.e. CGS, SG, and VR) produces a positive change in behaviour, knowledge or injury rate?

In step two, keywords were appended to guarantee that the criteria were maintained in the literature search: (construction) AND (safety training OR H&S training) AND (traditional OR passive OR education OR intervention); (construction) AND (safety training OR H&S training) AND (simulation); (construction) AND (safety training OR H&S training) AND (serious game OR game OR video game); (construction) AND (safety training OR H&S training) AND (virtual reality OR augmented reality OR mixed reality).

In step three, the search was undertaken on integrated databases to guarantee the interdisciplinary nature of this review. The databases included Web of Science, Taylor & Francis Online, Scopus, and ScienceDirect. Potentially eligible publications were retrieved (i.e. peer-reviewed journals, conference proceedings, technical and government reports and unpublished reports). In addition, publications recommended in database searches and referenced in the highest cited articles were also retrieved. The search yielded 569 hits on TT, 132 hits on CGS and SG, and 109 hits on VR. The 131 duplicates on TT, 26 duplicates on CGS and SG, and 17 duplicates on VR were removed (Figure 1).

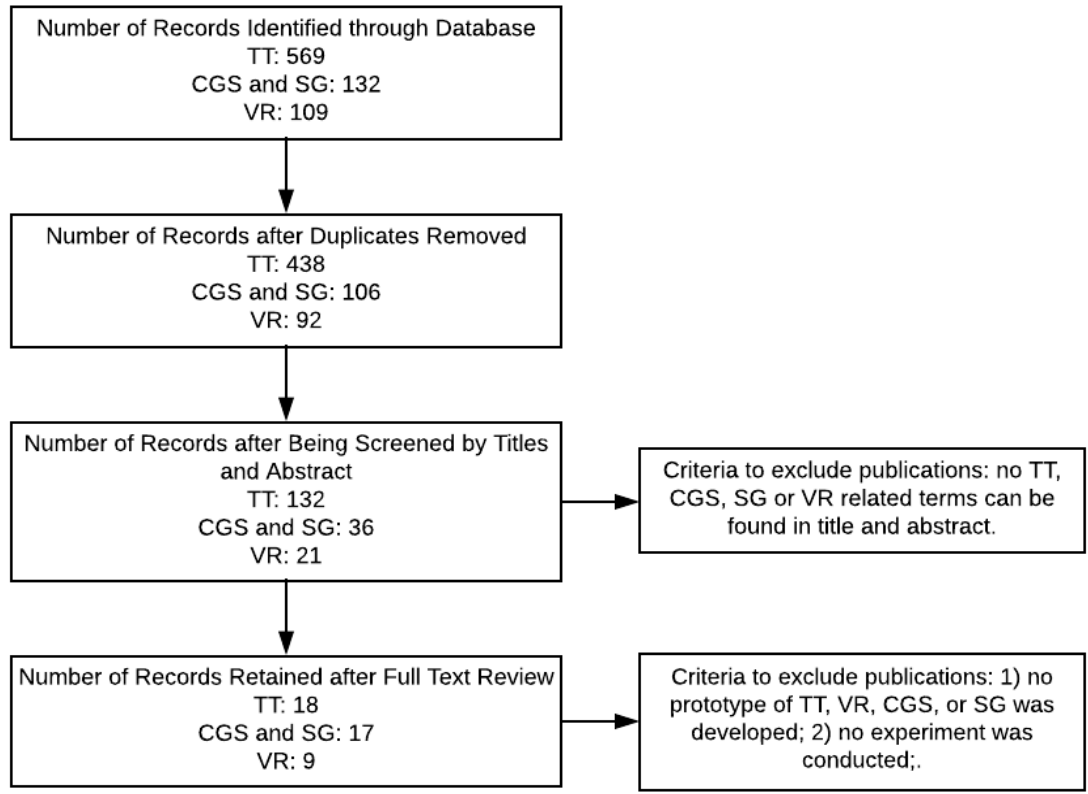

**Fig. 1.** Publication Filtering Process

In step four, a two-round examination was carried out to ensure academic rigor of the publication inclusion process. In the first round, titles and abstracts of retained publications were reviewed, and those had no TT, CGS, SG, and VR related terms were removed (Figure 1). In the second round, the full text of retained publications was assessed using the minimum acceptable criteria suggested by Feng, Gonzalez, Amor, Lovreglio, and Cabrera (2018): 1. a prototype of TT, VR, CGS, or SG should be developed; and 2. an experiment should be conducted to implement the prototype. Eventually, 18 articles on TT, 17 articles on CGS and SG, and nine articles on VR were retained (Figure 1).

In step five the evidence was summarised, and in step six the interpreted findings are presented in the following 'Results' section.

## 3. Results

### 3.1 The Use of Traditional Tools (TT)

#### 3.1.1 General Findings

In this section, a review of existing literature on the use of TT for construction H&S training is summarised to answer the first question predefined for this research: *is there enough evidence available to justify that training using TT makes a positive change in behaviour, knowledge or injury rate?* A total of 18 articles published between the year of 1986 and 2016 from 12 different journals were retrieved and examined on different aspects such as sample size, data collection method, factor to assess, and effectiveness. The results are summarised in Table 1 and Table 2.

**Table 1**

Authors, and Journal Names of Selected Studies of Traditional Tools

| Authors | Journal Names |
| --- | --- |
| (Chaffin, Gallay, Woolley, & Kuciemba, 1986) | International Journal of Industrial Ergonomics |
| (Saarela, 1989) | Journal of Safety Research |
| (Albers, Li, Lemasters, Sprague, Stinson, & Bhattacharya, 1997) | American Journal of Industrial Medicine |
| (Lusk, Hong, Ronis, Eakin, Kerr, & Early, 1999) | The Journal of the Human Factors and Ergonomics Society |
| (Johnson & Ruppe, 2002) | International Journal of Occupational Safety and Ergonomics |
| (Lingard, 2002) | Journal of Safety Research |
| (Spangenberg, Mikkelsen, Kines, Dyreborg, & Baarts, 2002) | Safety Science |
| (Gilkey, Hautaluoma, Ahmed, Keefe, Herron, & Bigelow, 2003) | American Industrial Hygiene Association Journal |
| (Darragh, Stallones, Bigelow, & Keefe, 2004) | American Journal of Industrial Medicine |
| (Hong, Ronis, Lusk, & Kee, 2006) | International Journal of Behavioural Medicine |
| (Kerr, Savik, Monsen, & Lusk, 2007) | Canadian Journal of Nursing Research |
| (Neitzel, Meischke, Daniell, Trabeau, Somers, & Seixas, 2008) | American Journal of Industrial Medicine |
| (Bena, Berchialla, Coffano, Debernardi, & Icardi, 2009) | American Journal of Industrial Medicine |
| (Sokas, Emile, Nickels, Gao, & Gittleman, 2009) | Public Health Reports |
| (Seixas et al., 2011) | International Journal of Audiology |
| (Adams et al., 2013) | Injury |
| (Forst et al., 2013) | American Journal of Industrial Medicine |
| (Evanoff et al., 2016) | Safety Science |

**Table 2**

Authors, Sample Sizes, Data Collection Methods, Assessment Factors, Retrieved Data (Pre, Post, Control, Treatment), and Effectiveness of Selected Studies of Traditional Tools

| Authors | Sample Sizes | Data Collection Methods | Assessment Factors | Pre-Training | Post-Training | Control Group | Treatment Group | Effectiveness |
|---|---|---|---|---|---|---|---|---|
| (Chaffin et al., 1986) | 33 | Supervisor-report (Observation) | Behaviour Alteration | Torso erect: 86% <br> Load close to body: 86% <br> Twist while lifting: 48% <br> Jerk load: 21% <br> Inadequate grips: 11% | Torso erect: 85% <br> Load close to body: 84% <br> Twist while lifting: 43% <br> Jerk load: 2% <br> Inadequate grips: 1% | | | 6.2% (on average) |
| (Saarela, 1989) | 60 | Supervisor-report (Injury Register) | Injury Rate Reduction | | 6% reduction (1 year after) | | | 6% |
| (Albers et al., 1997) | 37 | Self-report (Questionnaire) | Knowledge Acquisition | | | Class 1# Score: 68% <br> Class 2# Score: 75% <br> Class 3# Score: 51% <br> Class 4# Score: 74% | Class 1# Score: 85% <br> Class 2# Score: 80% <br> Class 3# Score: 58% <br> Class 4# Score: 86% | 10% (on average) |
| (Lusk et al., 1999) | 837 | Self-report (Questionnaire) | Behaviour Alteration | | | Group with training: 46% | Group without training: 52% | 6% |
| (Johnson & Ruppe, 2002) | 50 | Supervisor-report (Injury Register) | Injury Rate Reduction | 8 injuries in 59,600 hours | 8 injuries in 59,600 hours (1$^{st}$ year) <br> 6 injuries in 85,000 hours (2$^{nd}$ year) | | | 56.7% (1$^{st}$ year) <br> 47.0% (2$^{nd}$ year) |
| (Lingard, 2002) | 25 | Supervisor-report (Observation) | Behaviour Alteration | Use of tools: 94% <br> Access to heights: 51% <br> Use of PPE: 65% <br> Manual Handling: 85% | Use of tools: 98% <br> Access to heights: 93% <br> Use of PPE: 96% <br> Manual Handling: 80% | | | 18% |
| (Spangenberg et al., 2002) | Not Mentioned | Supervisor-report (Injury Register) | Injury Rate Reduction | | 25% reduction | | | 25% |
| (Gilkey et al., 2003) | 107 | Self-report (Questionnaire) | Behaviour Alteration | 71.8% | 76.8% | | | 5% |
| (Darragh et al., 2004) | Not Mentioned | Supervisor-report (Injury Register) | Injury Rate Reduction | 1,478 injuries in 16,946,918 hours | 493 injuries in 6,706,046 hours | | | 15.5% |
| (Hong et al., 2006) | 612 | Self-report (Questionnaire) | Behaviour Alteration | 50% | 57% | | | 7% |
| (Kerr et al., 2007) | 343 | Self-report (Questionnaire) | Behaviour Alteration | 42% | 50% | | | 8% |
| (Neitzel et al., 2008) | 23 | Self-report (Questionnaire) | Behaviour Alteration | 29.2% | 57.1% | | | 27.9% |
| (Bena et al., 2009) | 2795 | Supervisor-report (Injury Register) | Injury Rate Reduction | | 6% reduction | | | 6% |
| (Sokas et al., 2009) | 175 | Self-report (Questionnaire) | Knowledge Acquisition | Test Score <br> Fall Hazard Knowledge: 2.4 <br> Electronic Hazard Knowledge: 3.7 | Test Score <br> Fall Hazard Knowledge: 3.1 <br> Electronic Hazard Knowledge: 4.4 | | | 24.5% (on average) |
| (Seixas et al., 2011) | 176 | Self-report (Questionnaire) | Behaviour Alteration | Use of HPD: 34.5% | Use of HPD: 46.6% (immediate) <br> Use of HPD: 42.0% (2 months later) | | | 12.1% (immediate) <br> 7.5% (2 months later) |
| (Adams et al., 2013) | 204 | Supervisor-report (Injury Register) | Injury Rate Reduction | | | Standard Training <br> 13% reduction (3 months) <br> 7% reduction (6 months) | Enhanced Training <br> 16% reduction (3 months) <br> 12% reduction (6 months) | 16%/13% (3 months) <br> 12%/7% (6 months) |
| (Forst et al., 2013) | 446 | Self-report (Questionnaire) | Knowledge Acquisition | Electrical Score: 36.9% <br> Fall Score: 70% <br> Electrical Injury Score: 54.1% | Electrical Score: 55.3% <br> Fall Score: 72.9% <br> Electrical Injury Score: 58.6% | | | 8.6% (on average) |
| (Evanoff et al., 2016) | 1273 | Supervisor-report (Injury Register) | Injury Rate Reduction | 18.2 per 100 person-years | 14.5 per 100 person-years | | | 20.3% |

*Notes:* PPE – Personal Protective Equipment; HPD – Hearing Protection Devices.

As can be seen from Table 2, almost all of the selected studies indicated the size of the samples included in their experiments, excepting two studies in which no such information was provided (Darragh et al., 2004; Spangenberg et al., 2002). The sample size varied greatly among the selected studies, ranging from 23 (Neitzel et al., 2008) to 2795 workers (Bena et al., 2009) (Column 'Sample Sizes', Table 2). Data from participants can be gathered by means of two methods, namely, self-report, and supervisor-report (Probst, 2004). In the use of self-report, data is reported directly by participants based on their personal experiences (Northrup, 1997). In the use of supervisor-report, data is reported by managers or supervisors based on their objective observations of participants (Lusk, Baer, & Ronis, 1995). As can be seen in Column 'Data Collection Methods' in Table 2, nine studies used the self-report method whilst the other nine studies adopted the supervisor-report method. The use of self-report is appropriate for studies where a large sample size is considered (Patel & Jha, 2014). However, participants may not take a survey seriously. Insincere or dishonest reporting might impact the reliability of data and should be considered as a source of error in the experimental design (Lilienfeld & Fowler, 2006). Reliability can be better guaranteed with the supervisor-report method where data are collected in a more objective manner (Hazucha, Hezlett, & Schneider, 1993).

Effectiveness is essential in evaluating training approaches (Ho & Dzeng, 2010). According to the literature, the effectiveness of a training approach is determined by three major factors: knowledge acquisition, behaviour alteration, and injury rate reduction (Column 'Assessment Factors', Table 2). To evaluate effectiveness, different methods were used in the selected studies, namely, the pre-post analysis (e.g., Kerr et al., 2007) (Columns 'Pre-Training' and 'Post-Training', Table 2), and the control-treatment contrast analysis (e.g., Lusk et al., 1999) (Columns 'Control Group' and 'Treatment Group', Table 2). The pre-post analysis is a fundamental type of research experiment, and it evaluates the effectiveness of an approach by comparing the differences on knowledge acquisition, behaviour alteration and injury rate reduction that participants exhibit before and after the implementation of the approach (Montgomery, 2017). The control-treatment contrast analysis is another type of research experiment that examines effectiveness by comparing the differences on knowledge acquisition, behaviour alteration and injury rate reduction between control and treatment groups after the implementation of an approach (Montgomery, 2017). In particular, a control group, which is commonly used as the baseline to be contrasted, can be seen as a group of individuals that do not receive the standard training received by individuals from the treatment group in an experiment (Bailey, 2008). As can be seen, 15 studies used the pre-post analysis method to evaluate the effectiveness (Columns 'Pre-Training' and 'Post-Training', Table 2) whilst the other three studies adopted the control-treatment contrast analysis method (Columns 'Control Group' and 'Treatment Group', Table 2).

### 3.1.2 Effectiveness

A group of secondary data from Table 2 were retrieved from the selected studies. Nine studies indicated a positive effect less than or equal to ten percent, four studies indicated a positive effect between ten and 20 percent, and five studies indicated a positive effect more than 20 percent (Column 'Effectiveness', Table 2). In addition, Table 2 also reveals that the effectiveness of an approach may decrease over time. This can be explained by the reported long-term and short-term effectiveness in three studies (Adams et al., 2013; Johnson & Ruppe, 2002; Seixas et al., 2011) (Column 'Effectiveness', Table 2). The main difference between short-term and long-term is the time delay between the trial and measuring the effectiveness of the training. The short-term is often measured immediately after the completion of a training programme, while the long-term could be measured several weeks, months or years after the completion of a training programme. As can be seen in Column 'Effectiveness' in Table 2, the long-term effectiveness is shown to be smaller than the short-term one. It is pointed out by Zohar and Erev (2006) that a long-term effectiveness may be less directly tied to the training compared to the short-term but can reflect retaining effectiveness across time. If this is the case, discontinuing training may lead to a return of any observed improvement to the baseline level.

## 3.2 The Use of Computer-Aided Technologies (CAT)

### 3.2.1 General Findings

In this section, a review of existing literature on the use of CAT for construction H&S training is summarised to answer the second question predefined for this research: *is there enough evidence available to justify that training with CAT (i.e., CGS, SG, and VR) makes a positive change in behaviour, knowledge or injury rate?* A total of 26 articles published between the years 2009 and 2018 from nine journals and 12 conference proceedings were retrieved, including 17 articles on CGS and SG and nine articles on VR (Table 3). The selected studies were then examined on different aspects, namely, genre, entertaining effect, type of play environment, software, sample and sample size, and effectiveness. The data is summarised in Table 3.

Table 3

Authors, Journals/Conference Names, Genres, Entertaining Effects, Operating Environment, Software, Samples (Sizes), and Effectiveness of Selected Studies of Computer-Aided Technologies

| Authors | Journals/Conference Names | Genres | Entertaining Effects | Play Environment | Software | Samples (Sizes) | Effectiveness |
|---|---|---|---|---|---|---|---|
| (Zhao, Lucas, & Thabet, 2009) | Proceedings of the 2009 Winter Simulation Conference | SG | Health Point | Scenario-led | Torque 3D | | |
| (Ku & Mahabaleshwarkar, 2011) | Journal of Information Technology in Construction | SG | Role-playing | Sandbox | Second Life | | |
| (Lin, Son, & Rojas, 2011) | Journal of Information Technology in Construction | SG | Reward System | Sandbox | Torque 3D | Student (5) | |
| (Liaw, Lin, Li, & Chi, 2012) | Proceedings of the 2012 Construction Research Congress | SG | Reward System | Sandbox | Torque 3D | Student (39) | |
| (Dickinson, Woodard, Canas, Ahamed, & Lockston, 2011) | Journal of Information Technology in Construction | CGS | No | Scenario-led | XNA Game Studio | Student (57) | |
| (Li et al., 2012) | Automation in Construction | SG | Role-playing | Scenario-led | Unity3D | Worker (25) | |
| (Zhao, Thabet, McCoy, & Kleiner, 2012) | eWork and eBusiness in Architecture, Engineering and Construction 2012 | CGS | No | Scenario-led | Unity3D | | |
| (Le & Park, 2012) | Proceedings of IEEE International Conference on Teaching, Assessment, and Learning for Engineering | SG | Role-playing | Sandbox | Second Life | | |
| (Guo et al., 2012) | Accident Analysis and Prevention | SG | Role-playing | Scenario-led | 3DVIA | Worker (15) | |
| (Greuter & Tepe, 2013) | 6th Digital Games Research Association (DiGRA) Conference | SG | Health Point | Scenario-led | Unity3D | Student (27) | |
| (Newton, Lowe, Kember, Wang, & Davey, 2013) | Proceedings of the 13th International Conference on Construction Applications of Virtual Reality | SG | Health Point | Sandbox | CryEngine | | |
| (Leong, Goh, & Ieee, 2013) | 2013 IEEE International Games Innovation Conference | CGS | No | Sandbox | Unity3D | | |
| (Dawood, Miller, Patacas, & Kassem, 2014) | Journal of Information Technology in Construction | CGS | No | Sandbox | OpenSim | Student (12) | |
| (Fang, Teizer, & Marks, 2014) | Proceedings of the 2014 Construction Research Congress | CGS | No | Scenario-led | Not Mentioned | | |
| (Li, Lu, Chan, & Skitmore, 2015) | Automation in Construction | CGS | No | Scenario-led | Unity3D | | |
| (Zhao & Lucas, 2015) | International Journal of Injury Control and Safety Promotion | SG | Health Point | Scenario-led | Torque 3D | | |
| (Sacks, Perlman, & Barak, 2013) | Construction Management and Economics | VR | No | Scenario-led | EON Studio | Student (66) | 35.3/14.3 |
| (Bosché et al., 2015) | Journal of Computing in Civil Engineering | VR | No | Scenario-led | Unity3D | | |
| (Carozza, Bosché, & Abdel-Wahab, 2015) | Proceedings of the 15th International Conference on Construction Applications of Virtual Reality | VR | No | Scenario-led | Unity3D | | |
| (Pedro, Le, & Park, 2015) | Journal of Professional Issues in Engineering Education and Practice | CGS | No | Scenario-led | Blender | Student (25) | |
| (Hilfert, Teizer, & König, 2016) | Proceedings of the International Symposium on Automation and Robotics in Construction | VR | No | Scenario-led | Unreal Engine | | |
| (Jeelani, Han, & Albert, 2017) | Computing in Civil Engineering 2017 | VR | No | Sandbox | Unity3D | Student (4) | |
| (Hou, Chi, Tarng, Chai, Panuwatwanich, & Wang, 2017) | Automation in Construction | VR | No | Sandbox | Unity3D | Worker (20) | |
| (Azhar, 2017) | Procedia engineering | VR | No | Sandbox | Unity3D | | |
| (Pena & Ragan, 2017) | IEEE Virtual Reality 2017 | VR | No | Sandbox | Unity3D | Student (5) | |
| (Shamsudin, Mahmood, Rahim, Mohamad, & Masrom, 2018) | Advanced Science Letters | VR | No | Sandbox | Not Mentioned | | |

Notes: Data in column 'Effectiveness' are presented in the percentage format (100%).

Genre refers to the type of CAT used in a study, such as CGS, SG, and VR. In the selected studies, CGS and SG often share a blurred boundary for two reasons: 1. they are all simulations showing textures and models from the description of each object; and 2. the ability of text to give a full and accurate picture of CGS and SG is often limited. To distinguish SG from CGS, a key criterion was used: whether a simulation is composed of entertaining effects, such as health point, reward system, and role-playing. Health point refers to a built-in indicative bar assigned to the protagonist in a video game that can notify the player of the value of their immediate physical status (Zhao et al., 2009). Normally, one player starts with a life bar of 100 percent to indicate the full ability to function, and their follow-up choices in the game will constantly determine the fluctuation of their physical status: if a safe behaviour is taken, the health value will increase; if a risky behaviour is adopted, the health value will drop until reaching zero (Moore, 2016). Reward system refers to the component of a video game that offers incentives such as health, cash and equipment power-ups to motivate players to continue with the gameplay (Gao et al., 2017). Role-playing refers to the technique that provides players with the opportunity to assume personal attributes of the virtual character in a game and explore from a first-person perspective (Apperley, 2006). By using these criteria, 10 out 17 studies on SG and CGS were attributed to SG (e.g., Zhao et al., 2009), whilst the other seven studies were associated with CGS (e.g., Dickinson et al., 2011) (Columns 'Genres' and 'Entertaining Effects', Table 3).

There are two types of play environment that can be found in CAT approaches from the selected studies: *scenario-led* and *sandbox*. Scenario-led is considered as the progression-type of play in which limitations are placed to divide the environment into discrete sections to let the user have access to one chapter of the environment at a time; successfully completing the tasks in one chapter unlocks other segments with preconceived options (Dawood et al., 2014). Sandbox is an unstructured-type of play environment that allows the user to have full access to the complete environment and wander without constraints, choose tasks to interact with, and compose plans to achieve in-scene goals in a number of possible ways; there is no specified order in which the goals need to be achieved (Dawood et al., 2014). As can be seen in Column 'Play Environment' in Table 3, 14 of previous efforts are built in scenario-led (e.g., Guo et al., 2012), while the other 12 studies are based on the sandbox approach (e.g., Newton et al., 2013).

The application of CAT in the selected studies were able to simulate different hazards that could be seen on construction sites, including: acute risks (e.g., fall from height, exposure to electricity, and struck by objects), catastrophic risks (e.g., building collapse and excavation cave-ins), and health risks (e.g., uncontained asbestos, inappropriately stored chemicals, and wood dust) (e.g., Dawood et al., 2014; Le & Park, 2012). The interaction with simulated hazards was developed in various ways from just risk identification (e.g., Lin et al., 2011) to risk control (e.g., Greuter & Tepe, 2013). In the 'risk identification' studies, users were trained to explore the virtual environment and spot hazards (e.g., a hammer being placed on the edge of a scaffold). In the 'risk control' studies, users were taught to apply appropriate resources to mitigate pre-populated hazards (e.g., a number of barriers to be placed to prevent workers entering in a working area where heavy machinery was operating).

There are several software packages that can provide the capabilities to develop simulations as well as video games for risk identification and risk control. Game engines, which are the software designed for creators to develop video games, can be used as one type of the dedicated software (Mól, Jorge, & Couto, 2008; Ward, 2008). There are several options when it comes to choosing a game engine, and the mainstream ones are Unity3D, Torque3D, Unreal Engine and CryEngine (Zielke, 2010) (Column 'Software', Table 3). These four engines are extremely powerful and can support development for different platforms. Users can build their project and deploy to over 25 platforms with Unity3D: mobile (i.e., iOS, Android, Fire OS), VR (i.e., Oculus Rift, Google Cardboard Android & iOS, Steam VR PC/Mac, PlayStation VR, Gear VR, Daydream), augmented reality (i.e., Apple ARKit, Google ARCore, Windows Mixed Reality-Microsoft HoloLens, Vuforia), desktop (i.e., Windows, Universal Windows Platform, Mac, Linux/Steam OS, WebGL, Facebook Gameroom), console (i.e.,

PlayStation 4, PlayStation Vita, Xbox One, Wii U, Nintendo 3DS, Nintendo Switch), and TV (i.e., Android TV, Samsung SMART TV, tvOS). In contrast, Torque3D, Unreal Engine and CryEngine can only support a relatively limited number of platforms from the list above. The scripting languages are C++ for Torque3D, Unreal Engine and CryEngine users and C# for Unity3D users. It is pointed out that C# is certainly easier for beginners to learn compared to C++ (Chakravathi, 2016). By September 2015, Unity3D had become the development platform with the largest market share and best Asset Store with most free resources and tutorials uploaded by subscribers (Unity3D, 2015). Unity3D is also recommended by Lv, Tek, Da Silva, Empereur-Mot, Chavent, and Baaden (2013), Schranz and Rinner (2014), and Cristie, Berger, Bus, Kumar, and Klein (2015) to be more beginner-friendly compared with Torque3D, Unreal Engine and CryEngine.

### 3.2.2 Effectiveness

As mentioned, effectiveness is the magnitude of a training approach to help workers pick up H&S knowledge, alter unsafe behaviours and reduce injuries. Relevant data retrieved from the selected studies are presented in Columns 'Samples (Sizes)' and 'Effectiveness' in Table 3. As can be seen, 12 out of 26 studies provided valid information for their experiments: nine studies tested their approaches with students, and three studies tested their approaches with workers. However, only one out of these 12 studies examined the effectiveness (Sacks et al., 2013), while the other 11 studies evaluated their approaches mainly from the usability perspective (e.g., ease of use, interactivity, clarity of instructions, degree of engagement) and data on effectiveness were not available (e.g., Dawood et al., 2014; Pedro et al., 2015). The study by Sacks et al. (2013) demonstrated a significant increase (35.3%) of H&S knowledge acquisition from the VR training group (before: test score = 9.67, $p < 0.05$; after: test score = 13.08, $p < 0.05$) compared to a 14.3% increase from the traditional training group (before: test score = 9.77, $p < 0.05$; after: test score = 11.17, $p < 0.05$) (Column 'Effectiveness', Table 3). However, the data retrieved from Sacks et al. (2013) has limited validity because it was tested with a very specific sample (college students) rather than the expected target (construction workers).

## 4. Discussion

The use of CAT for H&S training had evidence supporting its effectiveness, but more solid evidence is yet required to support this statement. Therefore, it is reasonable to ask why many scholars were convinced of the power of CAT (e.g., Dickinson et al., 2011; Guo et al., 2012; Li et al., 2012; Zhao & Lucas, 2015). This section attempts to addresses this question by comparing the use of TT and CAT from several technical aspects. In the literature, the use of TT for H&S training is likely to have three major limitations, namely, limited representation of the actual workplace situations (Choudhry & Fang, 2008), limited consideration for workers who have low English proficiency (LEP) and low literacy (LL) (Choudhry & Fang, 2008), and failing to attract and maintain trainees' attention (Cherrett, Wills, Price, Maynard, & Dror, 2009). Instead, it is perceived that CAT can overcome the TT's limitations and bring economic benefits. More details of all these aspects are interpreted in the following section.

### 4.1 Representing the Actual Workplace Situations

TT training comprises both on-site and off-site programmes (Guo et al., 2017). On-site programmes directly expose trainees to real hazards and interfere with the construction progress by allowing trainees develop hands-on practices with workplace resources such as hand tools and machinery (Guo et al., 2012; Li et al., 2015). To overcome these issues, managers would rather arrange off-site training: workers are trained by attending workshops and memorising safety-related material written using technical jargon (Guo et al., 2012). So that workers can return to work as soon as possible, these training sessions are short duration (Guo et al., 2012). It is argued that knowledge learned in this way is impractical. To take an example from a Hong Kong-based study, an interviewed worker stated that TT training was a waste of time because it did not represent the actual workplace situations, and working on real construction sites was entirely different to how they were trained (Choudhry & Fang, 2008). Site conditions are complex and usually vary too fast with respect to weather, temperature, heat, humidity, and housekeeping for workers to apply theoretical H&S knowledge they have learned (Choudhry & Fang, 2008).

The expressive ability of TT is insufficient to fully showcase hazardous scenes to trainees (Li et al., 2012). Hazards can be better explained by using visual aids (Zhao & Lucas, 2015). Researchers are able to employ a range of techniques to give accurate representations of physical properties with CAT solutions (e.g., 3D Warehouse, Laser Scanner, BIM Toolkits, Autodesk Recap, and 360 Camera) (Höllerer, Feiner, Terauchi, Rashid, & Hallaway, 1999; Seth, Vance, & Oliver, 2011). Li et al. (2012) argue that the use of CAT allows simulation of actual site conditions by generating a close-to-reality virtual environment that contains "all available details, including both temporary and permanent structures, building services, construction material storage, waste, construction equipment and tools" (p. 501).

## 4.2 Text-free Interfaces

Loosemore and Andonakis (2007) point out that the construction sector workforce tends to have on average a lower educational attainment compared to other industries. In the US, according to an industrial survey by National Safety Council in 2003, more than 70% of the construction workforce had LEP issues (Vazquez & Stalnaker, 2004). In NZ, 7,485 Asian workers (who might have LEP and LL issues) representing almost 5% of the whole workforce, were employed in the construction sector in 2013 (NZ.Stat, 2013). TT approaches rely heavily on paper-based materials and textual descriptions and therefore require trainees to be literate; having LEP and LL issues could lead to an individual having difficulties in acquiring H&S knowledge (Gao et al., 2017). For example, a study of construction workers in Hong Kong revealed that uneducated workers could not read safety materials and had difficulties reporting and communicating during safety meetings (Choudhry & Fang, 2008). This highlights the importance of developing fair training for LEP and LL workers. Further evidence appears in a CAT study for H&S training in this field: a cohort of 27 students for whom English was the second language responded positively that a gaming approach produced a better learning experience than attending lectures because few text-based materials were used (Greuter & Tepe, 2013). Consequently, the use of CAT can mitigate language issues for LEP and LL workers.

### 4.3 User Engagement

It is argued that TT approaches are not engaging and trainees' attention is poor at best (Cherrett et al., 2009; Harfield et al., 2007), whereas CAT approaches can immerse users by providing three major elements: presence, flow, and character identification (Bachen, Hernández-Ramos, Raphael, & Waldron, 2016; Girard, Ecalle, & Magnan, 2013; Sacks et al., 2013). Presence is the mental experience occurring when a person is psychologically involved in a virtual environment (Bachen et al., 2016). Flow is the psychological sensation of losing track of time experienced by individuals who are busy responding to challenges that largely examine their stored knowledge (Bachen et al., 2016). Character identification is the instinctive reaction of individuals to assume as personal attributes the identities and goals of avatars in the virtual environment (Bachen et al., 2016). In particular, the nature of presence can arouse a frightening degree of sense (e.g., 'it can happen to you'), which ensures that a participatory response in the virtual environment is as close as possible to a natural response in real-life situations (Sacks, Whyte, Swissa, Raviv, Zhou, & Shapira, 2015; Zhao & Lucas, 2015).

In terms of the degree of user engagement, the literature evidence from the construction sector was not able to fully support the advantages of CAT over TT for H&S training because TT and CAT studies were developed under different methodological, conceptual and practical assumptions. Another consideration is that the studies analysed are spread across 20 years, and VR and SG technologies were in their infancy 20 years ago. However, pertinent research was also considered from other fields of science and engineering to gather more conclusive evidence. A study by Chittaro and Buttussi (2015) in the aviation safety area provides the valid evidence that the treatment group received CAT training had better user engagement (M=5.04, SD=0.94) compared to the control group received TT training (M=3.41, SD=1.24).

### 4.4 Cost Efficient

Recent evidence indicates that the making of a CAT training programme requires capital investment mainly in the purchase of hardware, and other necessities such as game engines (e.g., Unity 3D), 3D models (e.g., building element, avatar, and vehicle) and 2D textures (e.g., weather and particle system). These costs make it particularly economical for developers to acquire the necessary resources online (Newton et al., 2013). For hardware purchase, funding is mostly spent on computers and VR headsets (Giuseppe & Wiederhold, 2015; Newton et al., 2013). The option to purchase a VR headset could be $449 for an Oculus Rift headset working with PCs or $10 for a Google Cardboard headset working with mobile devices (Howard, 2018). In addition, the pricing strategy for game engines can range from free to more than $100 USD per month. *Unity3D* requires a $125 USD monthly fee from professionals and studios to sustain ownership of a license. A free version is made available to beginners and students for non-commercial development with nearly the same functionality as the charged version, just excluding real-time shadows and post-processing render effects. *Unreal Engine* is free for all levels of creators but can cost a business for 5% of commercialisation revenue when exceeding $3000 USD. *Torque3D* and *CryEngine* are made free to all levels of creators no matter how much they earn.

When developed, CAT approaches can be distributed as either low-priced apps for mobiles or playable simulations for platforms to let employees have access to the training resources for repeated learning at any time and place (Lin, Migliaccio, Azari, Lee, & De La Llata, 2012). In this case, the cost for construction companies can be greatly reduced whilst still achieving maximal training results (Ho & Dzeng, 2010). However, TT requires fixed enrolment fees from trainees to attend each training program (Forst et al., 2013), and thus is considered not as cost-efficient as CAT.

5. Theoretical framework to Assess CAT

Based on the findings from the systematic review, existing CAT studies lack a comprehensive evaluation of effectiveness and usability. As a result, we propose a theoretical framework for a systematic evaluation of CAT training approaches using the systematic review theoretical findings. This is shown in Figure 2. This framework may help future research to establish possible pathways to provide valid insights into the effectiveness and usability of their CAT training approaches. A pool of items suggested by the works of Lin et al. (2011) and Greuter and Tepe (2013) to evaluate the usability aspects are listed in Table 4. Note that we acknowledge that this framework is theoretical in nature and requires further validation. However, we believe it helps to establish a clear pathway for future research.

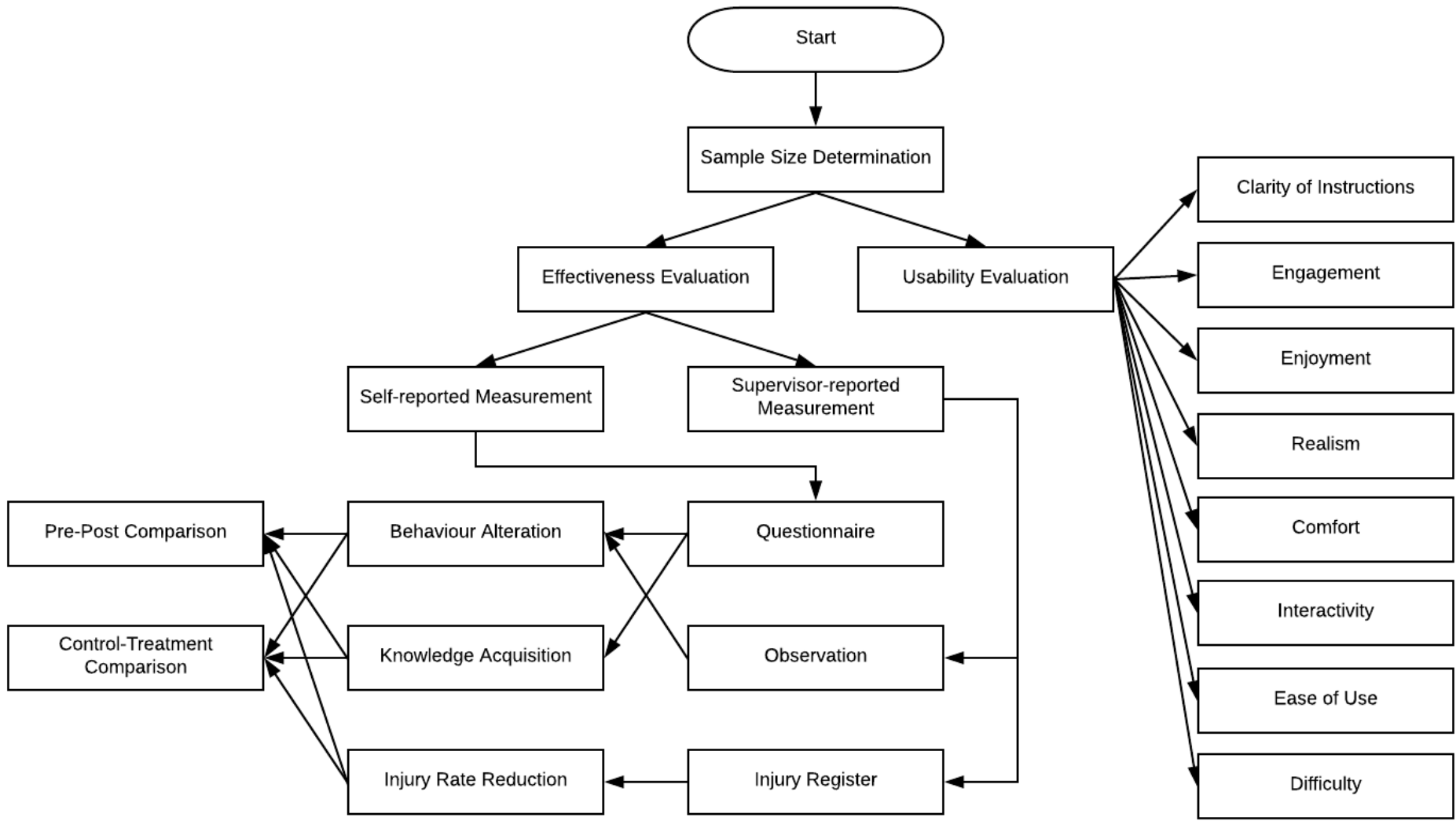

**Fig. 2.** A Theoretical Framework to Systematically Evaluate CAT Training Approaches

**Table 4**

Pool of Items to Evaluate the Usability Aspects of CAT Training Approaches

| Usability Aspects | Questions |
|---|---|
| Clarity of Instructions | Q1: Is the information provided in the system clear? |
| Engagement | Q2: To what extent did the system hold your attention? |
| | Q3: Did you lose track of time while playing with the system? |
| | Q4: How aware were you of events occurring around you? |
| | Q5: Were you disappointed when the session was over? |
| | Q6: Did you feel bored when playing with the system? |
| Enjoyment | Q7: How enjoyable did you find the graphics? |
| | Q8: How enjoyable did you find the sound? |
| | Q9: How much did you enjoy the way you interacted with the environment? |
| Realism | Q10: How realistic did the game reflect the everyday construction practical experience? |
| Comfort | Q11: Did you experience motion sickness when interacting with the virtual screen of the system? |
| Interactivity | Q12: Did you feel comfortable when interacting with the virtual screen of the system? |
| Ease of Use | Q13: Did you find the devices of the system easy to operate? |
| | Q14: Did the interface at the bottom provide you with too many options? |
| | Q15: Did the user interface frustrate you? |
| Difficulty | Q16: To what extend did you find the content challenging? |
| | Q17: Did you want to give up at any time during the session |
| | Q18: How often did you make mistakes? |

## 6. Future Research Directions

The systematic review findings have shown that the use of CAT has the potential to enhance training experience. Thus, this review helps to identify the limitations of existing CAT approaches that in turn represent the directions for further research in this field.

In the construction sector, some workers are more likely to get injured than others (Smith et al., 2015). For example, in the US, the fatality rate in 1993 was 15 per 100,000 in construction workers aged 65 years or more and five per 100,000 in workers aged 25 to 34 years (NIOSH, 2014). This data also revealed that Hispanics had the highest fatality rate (34.8 per 100,000 employed persons) compared to African Americans (24 per 100,000 employed persons) and European Americans (10.6 per 100,000 employed persons) in the New Jersey construction sector from 1983 to 1989 (Sorock, Smith, & Goldoft, 1993). Thus, the risk of getting injured might vary from group to group, even from person to person (Dong & Platner, 2004). However, previous research in this field focused solely on developing training approaches for the general population, and does little to consider individual differences. The descriptions in the existing studies have portrayed that CAT approaches have in general a pre-determined and very fixed training pathway for each individual, going through the same learning contents and experience (e.g., Guo et al., 2012; Li et al., 2012; Lin et al., 2011). This may be valid for most workers but may not be for those workers who are more susceptible to workplace accidents. More robust design should be considered. Future research in this field should better understand individual needs when developing training methods.

Additionally, as mentioned, the proposed framework for a systematic evaluation of CAT training approaches is theoretical in nature and requires further validation. Thus, further study is needed to test and validate this framework.

## 7. Conclusion

This paper is an attempt to provide insights into the effectiveness of using TT and CAT for H&S training in the construction sector. A systematic review of existing publications in this field was carried out in March 2018. As a result, the findings indicate that TT had sufficient statistical evidence to support its effectiveness. CAT also had support for its effectiveness, but more solid evidence is yet required to support this statement. This is the main contribution of this paper as the effectiveness of TT and CAT was not uncovered in previous review studies in this field. A further discussion between the technical characteristics of TT and CAT leads us to think that CAT is more powerful than TT in terms of representing the actual workplace situations, providing text-free interfaces, having better user engagement, and being more cost-efficient. Finally, a theoretical framework is proposed as a potential solution to help future research to establish possible pathways to systematically examine the effectiveness and usability of their CAT approaches. A further research is proposed to test and validate this framework.

The shortcomings of previous studies are also identified. In previous studies, CAT solutions were designed in a pre-determined and very fixed pathway so that each trainee needed to go through the same learning contents and experience. This may not be appropriate for everyone, especially for accident-prone workers. More robust design is therefore needed. Future research in this field should better understand individual needs when develop training methods.